\documentclass[twocolumn,showpacs,superscriptaddress,amssymb,10pt,prl,floatfix,aps,longbibliography]{revtex4-1}

\usepackage{graphicx}
\usepackage{amsmath}
\usepackage{dcolumn}
\usepackage{bm}
\usepackage{epsfig}
\usepackage{color}
\usepackage{longtable}
\usepackage{leftidx,amsmath}
\usepackage{natbib} 
\usepackage[utf8]{inputenc}
\usepackage{braket}
\usepackage{tikz}
\usetikzlibrary{snakes}
\usepackage[utf8]{inputenc}
\usetikzlibrary{snakes}

\def\emc2{\ensuremath{~m_\text{e}c^2}}

\begin{document}
	\preprint{}
	%
%
%
%
	\title{All-order Coulomb corrections to Delbrück scattering\\ above the pair production threshold}
%
%
%
%

	\author{J.~Sommerfeldt}
	\email{j.sommerfeldt@tu-braunschweig.de}
	\affiliation{Physikalisch--Technische Bundesanstalt, D--38116 Braunschweig, Germany}
	\affiliation{Technische Universit\"at Braunschweig, D--38106 Braunschweig, Germany}
	
	\author{V.~A.~Yerokhin}
	\affiliation{Physikalisch--Technische Bundesanstalt, D--38116 Braunschweig, Germany}
    \affiliation{Max-Planck-Institut für Kernphysik, D--69117 Heidelberg, Germany}
    
    \author{Th.~Stöhlker}
    \affiliation{Helmholtz Institute Jena, D--07743 Jena, Germany}
    \affiliation{GSI Helmholtzzentrum für Schwerionenforschung GmbH, D--64291 Darmstadt, Germany}
	
	\author{A.~Surzhykov}
	\affiliation{Physikalisch--Technische Bundesanstalt, D--38116 Braunschweig, Germany}
	\affiliation{Technische Universit\"at Braunschweig, D--38106 Braunschweig, Germany}	

	\date{\today \\[0.3cm]}

%
%
\begin{abstract}
We report calculations of Delbrück scattering that include all-order Coulomb corrections for photon energies above the threshold of electron-positron pair creation. Our approach is based on the application of the Dirac-Coulomb Green function and accounts for the interaction between the virtual electron-positron pair and the nucleus to all orders in the nuclear binding strength parameter $\alpha Z$. Practical calculations are performed for the scattering of 2.754~MeV photons off plutonium atoms. We find that including the Coulomb corrections enhances the scattering cross section by up to 50\% in this case. The obtained results resolve the long-standing discrepancy between experimental data and theoretical predictions and demonstrate that an accurate treatment of the Coulomb corrections is crucial for the interpretation of existing and guidance of future Delbrück scattering experiments on heavy atoms.
\end{abstract}
\maketitle
\textit{Introduction.---}Delbrück scattering is the elastic scattering of photons by the Coulomb field of an atomic nucleus. This process is not allowed in linear electrodynamics but becomes possible in quantum electrodynamics (QED) via virtual electron-positron pair production. Delbrück scattering is one of the very few non-linear QED processes that can be precisely studied in experiment. Combined with accurate theoretical predictions, experimental observations of this fundamental process can provide stringent tests of QED theory and yield important information on the nuclear structure~\cite{meitner_uber_1933,SCHUMACHER1999101, MILSTEIN1994183}. So far, these possibilities have not been fully explored because a sufficiently accurate theoretical description of Delbrück scattering is not available. The most widely used approach in the region of moderate energies of 1-10~MeV is the lowest-order Born approximation as developed by Papatzacos and Mork~\cite{PhysRevD.12.206}. This approximation is based upon expanding the Delbrück amplitude in the Coulomb-field strength parameter $\alpha Z$, where $Z$ is the nuclear charge and $\alpha$ is the fine-structure constant, and neglecting all terms beyond the lowest order that are usually referred to as Coulomb corrections. For heavy nuclei, this approach is not applicable as the Coulomb corrections grow with the nuclear charge and are expected to drastically change the cross section compared to the Born approximation~\cite{SCHUMACHER1999101}. However, analysis of this high-$Z$ regime attracts particular attention in experiment and theory as a testbed to explore non-perturbative QED, which can not rely on $\alpha Z$ expansions. Moreover, the Delbrück cross section scales as $Z^4$, allowing experiments on high-$Z$ targets to achieve an increasingly high accuracy.

In order to better understand Delbrück scattering in the non-perturbative high-Z regime, new experiments are planned that use nuclear $\gamma$-sources, Compton scattering techniques, and novel accelerator facilities that will employ gamma rays with energies above the electron-positron pair production threshold~\cite{PhysRevSTAB.8.100702, PhysRevSTAB.17.033501, PhysRevLett.78.4569, PhysRevSTAB.14.050703, krasny_gamma_2015, budker_atomic_2020}. The necessary prerequisite for the success of these experiments is a breakthrough in the theory of Delbrück scattering, since the current theory is not adequate in this parameter region, as was repeatedly stressed in the literature~\cite{SCHUMACHER1999101,RULLHUSEN1979166,rullhusen_coulomb_1979, PhysRevC.23.1375, SCHUMACHER1975134}. In particular, the measured angle differential cross section for the scattering of 2.754~MeV photons off neutral plutonium atoms was found to differ from the lowest-order Born predictions by almost a factor of two~\cite{rullhusen_coulomb_1979}. This long-standing discrepancy has not been resolved up to now, despite the considerable interest it has attracted.

The first steps to set up an {\it ab initio} theory that accounts for all orders in the Coulomb-field strength parameter for Delbrück scattering were performed by Scherdin and co-workers~\cite{PhysRevD.45.2982,scherdin_coulomb_1995}. However, due to overwhelming technical difficulties, no actual calculations for energies above the pair-creation threshold were carried out. In this letter, we propose a novel approach to evaluate and compute Delbrück scattering amplitudes for a wide range of scattering energies including those beyond the threshold. This approach is based on the use of the Dirac-Coulomb Green function, which is exact in $\alpha Z$, and on a modified Wick-rotated integration contour. Our approach allows for accurate Delbrück calculations in the high-Z and high photon energy regime. To illustrate the application of the proposed theoretical method, we present calculations for the scattering of 2.754~MeV photons by plutonium atoms. We demonstrate that by accounting for the higher-order Coulomb corrections, one can resolve the long-standing discrepancy and reproduce the experimental results by Rullhusen and co-workers~\cite{rullhusen_coulomb_1979}. Relativistic units~(r.u.) {$\hbar = m_e = c = 1$} are used throughout this paper, if not stated otherwise.\\

\begin{figure}
\begin{center}
\includegraphics[scale=1.]{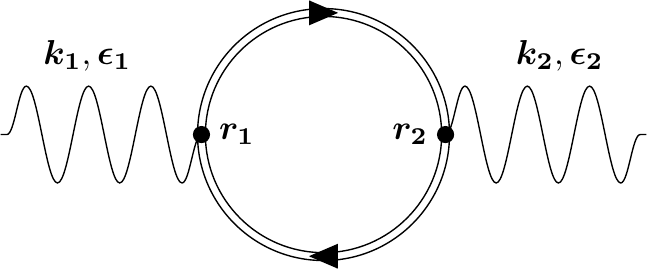}\caption{Feynman diagram for Delbrück scattering to leading order in $\alpha$ and all orders in $\alpha Z$.}\label{FeynmanDel}
\end{center}
\end{figure}

\textit{Theoretical Background.---}The Feynman diagram for Delbrück scattering is depicted in Fig.~\ref{FeynmanDel}. Here, we follow the standard convention where the wavy lines represent the incoming/outgoing photon with wave vector $\boldsymbol{k_1}/\boldsymbol{k_2}$ and polarization vector $\boldsymbol{\epsilon_1}/\boldsymbol{\epsilon_2}$, and the double lines correspond to the virtual electron-positron pair in the Coulomb field of the nucleus. Each vertex $\boldsymbol{r_1}$/$\boldsymbol{r_2}$ contributes a factor of $\sqrt{\alpha}$ to the scattering amplitude while the Dirac-Coulomb solutions for the electron and positron account for all orders in $\alpha Z$.

According to the well-known Feynman correspondence rules, the amplitude for the diagram in Fig.~\ref{FeynmanDel} can be written as

\begin{equation} \label{MatrixElement}
\begin{aligned}
M^D_{\epsilon_1,\epsilon_2} &= \frac{i\alpha}{2\pi} \int_{-\infty}^\infty \text{d}z~\int_{-\infty}^\infty \text{d}z'~\int \text{d}^3\boldsymbol{r}_1~\int \text{d}^3\boldsymbol{r}_2~\\
&\times\text{Tr}\Big[\hat{R}(\boldsymbol{r}_1,\boldsymbol{k}_1,\boldsymbol{\epsilon}_1) G(\boldsymbol{r}_1,\boldsymbol{r}_2,z)\hat{R}^\dagger(\boldsymbol{r}_2,\boldsymbol{k}_2,\boldsymbol{\epsilon}_2)\\
&\times G(\boldsymbol{r}_2,\boldsymbol{r}_1,z')\Big]\delta (\omega+z-z')~.\\
\end{aligned}
\end{equation}

\noindent Here, $G(\boldsymbol{r}_2,\boldsymbol{r}_1,z)$ is the Dirac-Coulomb Green function with the three-dimensional coordinate vectors $\boldsymbol{r}_1$ and $\boldsymbol{r}_2$ as well as the energy argument $z$. Moreover, $\hat{R}(\boldsymbol{r},\boldsymbol{k},\boldsymbol{\epsilon})$ is the photon-lepton interaction operator with $\boldsymbol{k}$ and $\boldsymbol{\epsilon}$ being the wave and polarization vectors and $\omega$ being the energy of the incoming and outgoing photon~\cite{MILSTEIN1994183}.

\begin{figure}
\includegraphics[scale=1.]{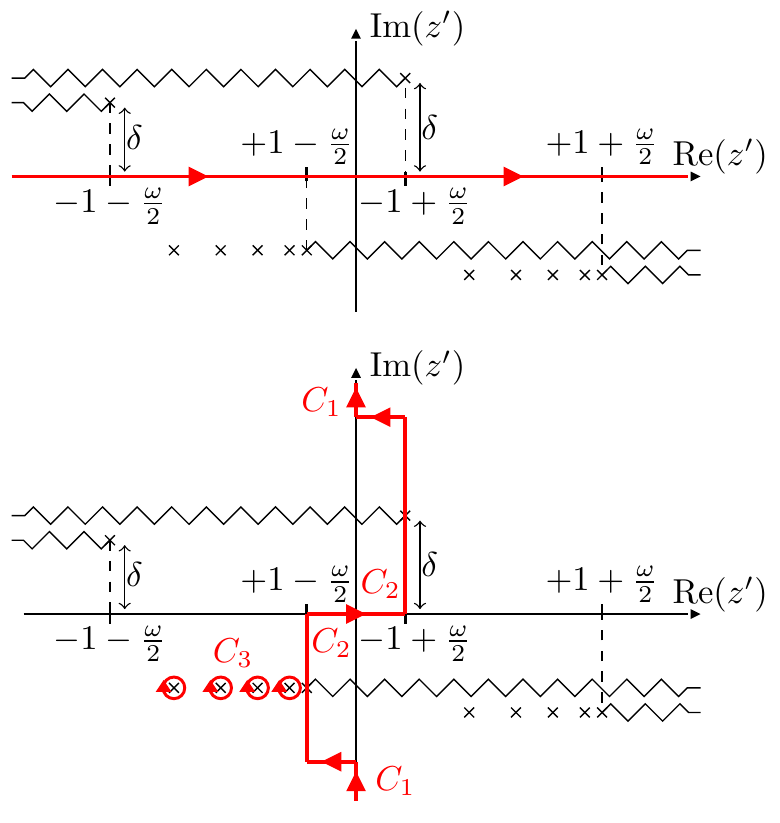}\caption{Original (upper panel) and Wick rotated (lower panel) contour for the $z'$ integration in the Delbrück amplitude~\eqref{MatrixElement}. The modified contour consists of paths along the imaginary axis ($C_1$) and around the branch cuts ($C_2$) as well as contributions from the residue ($C_3$). Moreover, the singularities (black crosses) and branch cuts (black zig zag lines) of the Green functions in Eq.~\eqref{MatrixElement} are shown for the case of above threshold photon energies.} \label{IntPath}
\end{figure}

Further evaluation of the amplitude~\eqref{MatrixElement} for the case when the photon energy is below the threshold for electron-positron pair production was discussed by us in Ref.~\cite{PhysRevA.105.022804}. However, the previous approach has to be extended for the more troublesome case of above threshold photon scattering. In what follows, therefore, we will discuss the challenges of this high energy analysis, $\omega \geq 2~\text{r.u.}$, and refer for all other details to Ref.~\cite{PhysRevA.105.022804}. First, to get sensible results from the Feynman diagram in Fig.~\ref{FeynmanDel}, one needs to eliminate the divergent free-loop contribution from the amplitude. As usual in bound-state QED calculations, the free diagram can be obtained by setting the nuclear charge to zero in the amplitude, see e.g.~\cite{MOHR1998227}. Now, we are ready to perform the energy integration in Eq.~\eqref{MatrixElement} over $z$ and $z'$. While the integral over $z$ is trivial due to the Dirac delta function $\delta(\omega+z-z')$, the substitution $z' \to z' + \tfrac{1}{2}$ simplifies the numerical $z'$-integration. Indeed, as seen from Fig.~\ref{IntPath}, the starting points of the branch cuts of the Green functions are located after this substitution at $z' = \pm 1 + \tfrac{\omega}{2}$ and $z' = \pm 1 - \tfrac{\omega}{2}$, i.e. \textit{symmetrically} with respect to the coordinate origin. Such symmetry makes calculations more stable as shown in Ref.~\cite{PhysRevA.105.022804}. Along with the cuts, one can also see two sets of bound-state poles localized at $z' = (\lambda' + m)/\sqrt{\gamma^2 + (\lambda'+m)^2} + \frac{\omega}{2}$ and $z' = (\lambda + m)/\sqrt{\gamma^2 + (\lambda+m)^2} - \frac{\omega}{2}$, $m = 0,1,2,...$. Both, the branch cuts and the two sets of poles, moreover, are shifted by $\delta \to \pm i0$ off the real axis. 

The infinitely close location of the poles of the Green function to the naive integration path on the interval $z' \in[-\infty,+\infty]$, displayed in the upper panel of Fig.~\ref{IntPath}, makes the numerical evaluation of the amplitude~\eqref{MatrixElement} very troublesome. It is more convenient to perform the well-known Wick rotation of the contour and integrate along the imaginary axis instead. However, in contrast to the previous below-threshold calculations, the branch cuts cross the imaginary axis for photon energies $\omega \geq 2~\text{r.u.}$ In order to overcome this difficulty, we follow Ref.~\cite{scherdin_coulomb_1995} and modify the Wick-rotated contour to incorporate additional paths that go around the branch cuts ($C_2$) and add the residue of the integrand for the enclosed poles ($C_3$), see lower panel of Fig.~\ref{IntPath}. By employing this modified contour, we finally obtain
\nopagebreak
\begin{equation} \label{intReplace}
\int_{-\infty}^{\infty} \text{d}z' f(z') = \int_{C_1,C_2} \text{d}z' f(z') - 2\pi i \sum_n \text{Res}[f,z'_n]~,
\end{equation}
\nopagebreak
\noindent where for the sake of brevity, we used the notation $f(z')$ for the integrand in Eq.~\eqref{MatrixElement} with $\text{Res}[f,z'_n]$ being the residue of $f(z')$ at its $n$th enclosed pole $z'_n$.

The evaluation of Eq.~\eqref{intReplace} requires the calculation of the integral along the paths $C_1$ and $C_2$ as well as the summation over the residue of the poles of the integrand $C_3$. The methods used to evaluate the path along the imaginary axis $C_1$ are identical to the case of below-threshold energies and are discussed in detail in Ref.~\cite{PhysRevA.105.022804}. In contrast, the integration along $C_2$ and the summation over the residue $C_3$ requires some important modifications that need to be discussed. For example, when integrating along $C_2$, we find that the integrand is strongly peaked at the beginning of the branch cuts at $z' = \pm 1 \mp \tfrac{\omega}{2}$. To perform the integration, we use Gauss-Legendre quadrature with an enhanced density of integration points close to the peaked regions which can be achieved by the substitution $z' \to \pm u^2 + (\pm 1 \mp \tfrac{\omega}{2})$. To calculate the residue $C_3$, we remind that the poles originate from a prefactor $\Gamma(\lambda'-\nu')$ arising in the radial components of the Green function~\cite{MOHR1998227}, where $\lambda' = \sqrt{\kappa'^2 - (\alpha Z)^2}$, $\nu' = \alpha Z (z'+\tfrac{\omega}{2})/c'$ and $c' = \sqrt{1-(z'+\tfrac{\omega}{2})^2}$. Therefore, to obtain the contribution from the bound states, we simply replace this prefactor by its residue
\nopagebreak
\begin{equation}
\begin{aligned}
\text{Res}\left[\Gamma(\lambda' - \nu'), z'_n = \frac{\lambda' + n}{\sqrt{\gamma^2 + (\lambda' + n)^2}} - \frac{\omega}{2}\right]\\ 
= -\frac{(-1)^n}{n!} \frac{\left(1- \frac{(\lambda' + n)^2}{\gamma^2 + (\lambda' + n)^2}\right)^{3/2}}{\gamma}~.
\end{aligned}
\end{equation}
\nopagebreak
Together with the integration over the energy $z'$, the evaluation of the radial integrals in Eq.~\eqref{MatrixElement} is also a highly demanding task. This is due to the fact that after some algebra presented in Ref.~\cite{PhysRevA.105.022804}, the integrand in Eq.~\eqref{MatrixElement} can be written in terms of products of Whittaker functions $M_{\alpha,\beta}(2cr)$ and $W_{\alpha,\beta}(2cr)$ which are fast oscillating and slowly decreasing at large radial arguments. To compute the integrals over $r_1$ and $r_2$, we split the integrals into two parts. For small radial distances, we perform the radial integrations numerically, whereas for large distances we employ the asymptotic expansion of the Whittaker functions to calculate the integrals analytically. In contrast to the below-threshold case of Ref.~\cite{PhysRevA.105.022804}, however, the analysis of the asymptotic representation of the Whittaker functions requires some special attention if $\omega \geq 2~\text{r.u.}$ The reason for this is the second term of the asymptotic expansion of $M_{\alpha,\beta}(2cr) \propto M_1 e^{cr} + M_2 e^{-cr}$ which is always exponentially smaller than the first term for $\omega < 2~\text{r.u.}$ but can be of comparable magnitude for higher energies. The radial integral including this term was also derived analytically in terms of incomplete Gamma functions. In general, the radial integration is a very time consuming task which was accomplished by utilizing a hybrid parallelization scheme  at the PTB high performance cluster. This allowed us to perform a full calculation for one charge number and photon energy in approximately  one week using 200 threads.

\begin{figure}
\includegraphics[scale=1.]{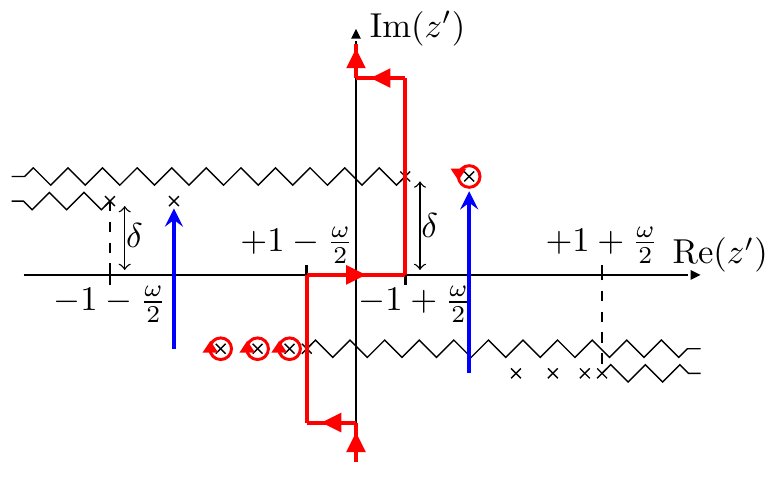}\caption{Analytic structure of the integrand and the integration contour after the redefinition of the vacuum to include the lowest-lying bound state. The corresponding pole in the lower left quadrant moves up and is not encircled anymore, whereas the pole in the lower right quadrant is also moving up and gets encircled by the integration contour.} \label{IntPathRV}
\end{figure}

So far, we have discussed the theory used to calculate Delbrück scattering amplitudes. With the help of these amplitudes, one can calculate the angle-differential as well as total cross sections of the Delbrück process. However, in order to compare the results of our calculations with experimental data, we have to account also for competing scattering processes. These are the Rayleigh scattering off bound atomic electrons and nuclear Thomson scattering. The Rayleigh and Delbrück scattering processes are closely related as can be readily seen in the redefined vacuum approach~\cite{shabaev_two-time_2002, sym13061014,PhysRevA.103.042818}. In this approach, the vacuum Fermi level in the Delbrück amplitude is shifted to include the Dirac energies of the occupied atomic shells. In practice, it implies changing the sign of the infinitesimal imaginary additions for the poles of the electron propagator corresponding to the occupied shells. The expressions for the Rayleigh amplitude are then obtained as the difference of the Delbrück amplitudes with the modified and the standard vacuum, as illustrated in Fig.~\ref{IntPathRV}. We have checked that formulas obtained in this way agree with the known expressions for Rayleigh scattering \cite{Surzhykov_2015}. We, therefore, employ our numerical procedure developed for Delbrück amplitudes to include Rayleigh scattering.

\begin{figure*}
    \centering
      \includegraphics[width=0.95\textwidth]{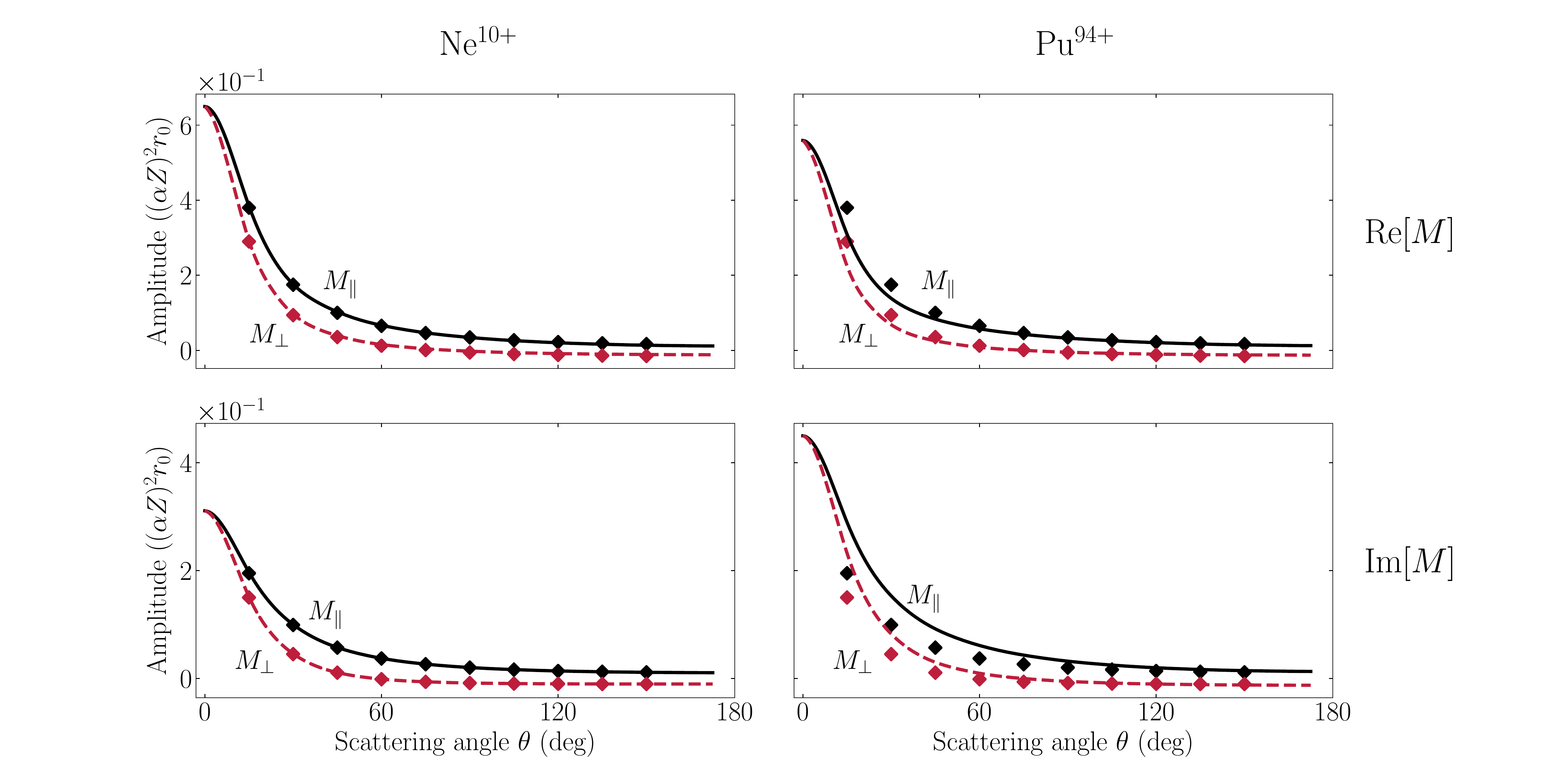}
       \caption{Real (upper panels) and imaginary (lower panels) parts of the amplitude for Delbrück scattering~\eqref{MatrixElement} of 2.754~MeV photons by bare neon (left panels) and plutonium (right panels) nuclei. Calculations have been performed for linear polarization of the incoming/outgoing photons parallel (black solid line) as well as perpendicular (red dashed line) to the scattering plane. Moreover, the lowest-order Born predictions from Ref.~\cite{FALKENBERG19921} are shown (diamonds). The amplitudes are given in units $(\alpha Z)^2r_0$, where $r_0 = 2.818~\text{fm}$ is the classical electron radius. \label{Amplitdues}}
\end{figure*}

In contrast to the Rayleigh and Delbrück processes that are closely related in the framework of QED, a simple approach can be used to describe the nuclear Thomson scattering. Namely, the amplitude for a rigid spin-zero nucleus with charge radius $R$ is given by
\nopagebreak
\begin{align}
M_{\perp}^T &= - \frac{\alpha Z^2}{M} \left(1 - \frac{1}{3} \omega^2 \langle R^2 \rangle \right) ~,\label{ThomsPerp}\\
M_{\parallel}^T &= M_{\perp}^T \cos \theta~, \label{ThomsPar}
\end{align}
\nopagebreak
\noindent where $Z$ and $M$ are the charge number and mass of the nucleus, see \cite{HUTT2000457, PhysRev.96.1428}. The two amplitudes~\eqref{ThomsPerp} and \eqref{ThomsPar} correspond to the scattering of photons that are linearly polarized within or perpendicular to the scattering plane spanned by the wave vectors $\boldsymbol{k_1}$ and $\boldsymbol{k_2}$.

\textit{Estimate of the theoretical uncertainty.---}To compare our theoretical results to experiment in a meaningful way, we need to estimate their uncertainty due to omitted higher-order effects. These effects are of several kinds: electron-electron interactions, QED contributions that are of higher-order in $\alpha$, Rayleigh scattering from higher-$l$ shells and nuclear structure effects. Since the electron-electron interaction corrections to the Delbrück process are difficult to estimate, we assume that they are of the same relative size as the ones for Rayleigh scattering. In the work by Volotka and co-workers~\cite{PhysRevA.93.023418}, it was shown that the inter-electronic interactions modify the differential cross section for Rayleigh scattering by about 2\% at 150 keV and become even smaller for higher energies. Although no estimate of QED corrections was made in Ref.~\cite{PhysRevA.93.023418}, they are suppressed by the small parameter $\alpha$. We thus conservatively estimate the combined uncertainty due to the electron-electron interaction and higher-order QED effects to be about 3\%. 

For the calculation of the Rayleigh scattering amplitudes, we account only for the K- and L-shells of the atom and neglect outer shell contributions. This approximation is justified by results presented in Ref.~\cite{PhysRevA.107.012805}, where it was shown that outer shells contribute remarkably only for small scattering angles and their role is reduced with increase of the energy. For the scattering angle of $45^\circ$ and incident photon energy of 175~keV, a 5\% contribution of outer shells was predicted in Ref.~\cite{PhysRevA.107.012805} and is taken as our error estimate. Yet another source of uncertainty is the contribution of the nuclear giant dipole resonance~(GDR) scattering. In the present letter we estimate this contribution from the existing data on the photo-nuclear absorption, as given by Eqs.~(3) and (4) of Ref.~\cite{rullhusen_coulomb_1979}. We estimate the corresponding uncertainty as 100\% of the resulting GDR contribution.

\textit{Results and discussion.---}We have discussed above the details of calculating elastic photon scattering amplitudes above the pair production threshold. Before employing these amplitudes to obtain cross sections relevant for $\gamma(2.754~\text{MeV})+\text{Pu}$ scattering experiments~\cite{rullhusen_coulomb_1979}, let us first examine the Delbrück case separately. In Fig.~\ref{Amplitdues}, we display the Delbrück amplitudes for collisions of 2.754~MeV photons with bare neon and plutonium nuclei. For each scenario, we present amplitudes for the scattering of photons that are linearly polarized either within or perpendicular to the plane spanned by $\boldsymbol{k_1}$ and $\boldsymbol{k_2}$. As it was shown in Ref.~\cite{KANE198675} based on symmetry considerations, all observables of the scattering process can be obtained from these two linearly independent amplitudes.

Apart of all-order in $\alpha Z$ calculations, the lowest-order Born predictions are also displayed in Fig.~\ref{Amplitdues}. The well established Born approximation~\cite{PhysRevD.12.206} is obtained by neglecting terms of order $(\alpha Z)^4$ and higher in the analysis of the Feynman diagram in Fig.~\ref{FeynmanDel}. As seen from the left panels of Fig.~\ref{Amplitdues}, the Born approximation and the all-order results agree very well for the case of bare neon. This is well expected for the low-Z regime where beyond-$(\alpha Z)^2$ terms are small. The higher-order corrections are enhanced, however, in the high-Z domain where they lead to remarkable modifications of the scattering amplitude. Indeed, as seen from the right panels of Fig.~\ref{Amplitdues}, the imaginary parts of $M_\parallel$ and $M_\perp$ are enhanced by about a factor of 1.6 and 2.8, respectively, for $\theta = 45^\circ$ if higher-order terms are taken into account. Such a paramount difference between Born approximation and all-order results is observed only for the energies above the threshold of pair production. For energies below the threshold, our calculations have shown that the Coulomb corrections do not exceed 15\%~\cite{PhysRevA.105.022804}.

\begin{figure}
    \centering
      \includegraphics[width=0.95\linewidth]{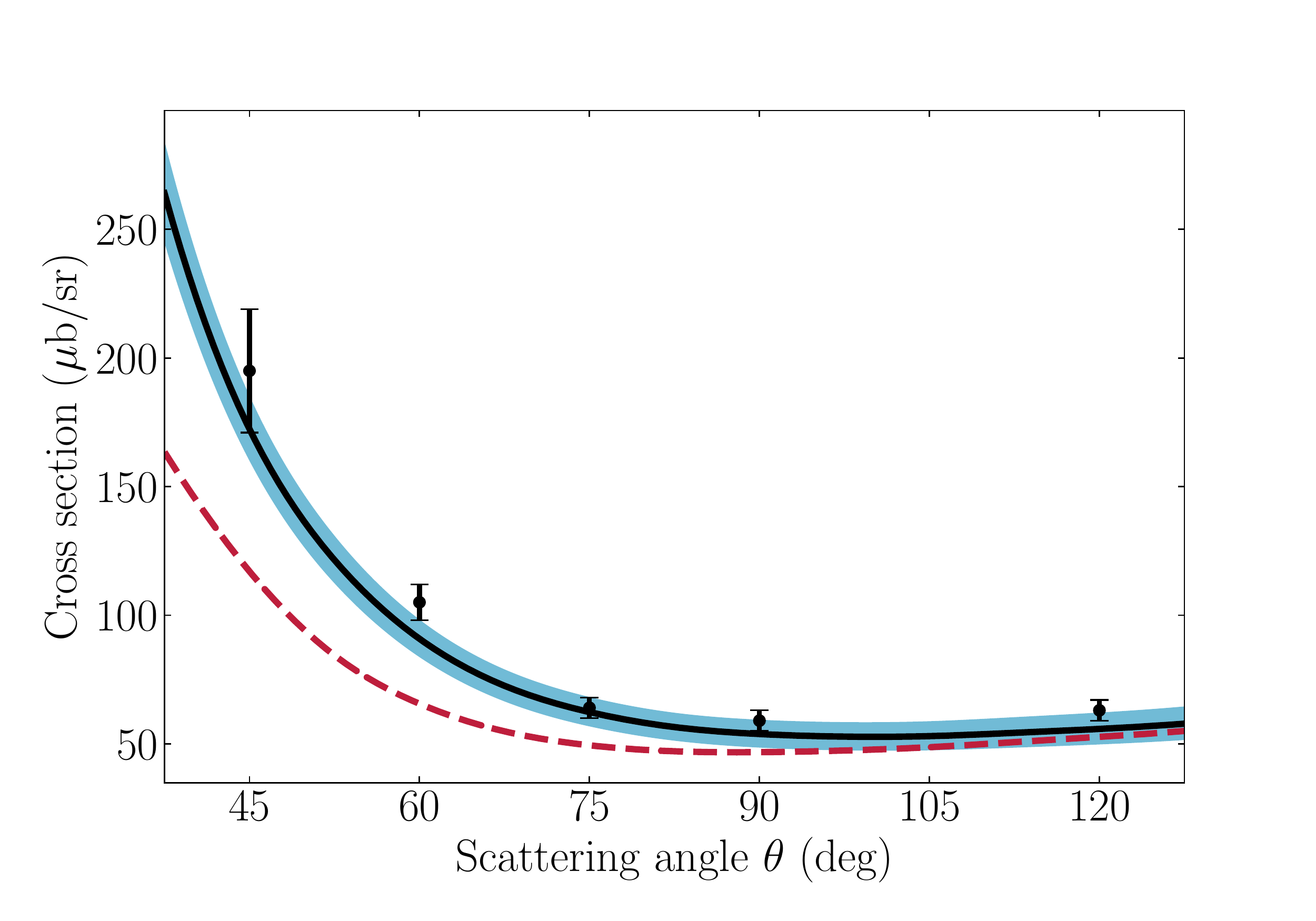}
       \caption{Differential cross section for elastic scattering of 2.754~MeV unpolarized photons by plutonium atoms. The black dots display the experimental data from Ref.~\cite{rullhusen_coulomb_1979}, the black solid line indicates the theoretical results based on all-order in $\alpha Z$ Delbrück calculations while the shaded region shows the theoretical error. Theoretical predictions using the lowest-order Born approximation for Delbrück scattering are displayed with the red dashed line. \label{CrossSections}}
\end{figure}

We are ready now to calculate the angle differential cross section of the elastic photon scattering by plutonium atoms. As shown in Ref.~\cite{KANE198675}, this cross section can be obtained as
\nopagebreak
\begin{equation}
\frac{\text{d}\sigma}{\text{d}\Omega} = \frac{1}{2}\left(\vert M_\parallel \vert ^2 + \vert M_\perp \vert ^2\right)~,
\end{equation}

\noindent where $M_{\parallel/\perp}$ is the sum of the amplitudes for Delbrück, Rayleigh as well as nuclear Thomson scattering and where the incoming radiation is assumed to be unpolarized. In Fig.~\ref{CrossSections}, we display this cross section together with its theoretical uncertainty and the experimental findings from Ref.~\cite{rullhusen_coulomb_1979}. Moreover, we present the theoretical predictions based on the lowest-order Born approximation for the Delbrück amplitude. As seen from the figure, the higher-order Coulomb corrections to the Delbrück process lead to a strong enhancement of the cross section for scattering angles $\theta < 90^\circ$. All-order in $\alpha Z$ predictions agree well with experimental data from Ref.~\cite{rullhusen_coulomb_1979}, thus, solving the long standing discrepancy between experiment and lowest-order Born theory. This agreement together with the computational stability of our analysis justifies the use of the proposed method for all-order calculations of Delbrück scattering for photon energies above the pair production threshold. In the future, such calculations will be performed to plan and to analyse Delbrück scattering experiments. These experiments are planned to be focussed not only on the total and differential cross sections but also on the polarization of the scattered photons which might be even more sensitive to higher-order Coulomb corrections.\\

{\begin{acknowledgments}
This work has been supported by the GSI Helmholtz Centre for Heavy Ion Research under the project BSSURZ1922. We thank our colleagues from PTB's high performance computing division and especially Gert Lindner for providing access to their computation cluster and for their excellent technical support. We also thank Sebastian Ulbricht and Sophia Strnat for very helpful discussions.
\end{acknowledgments}}


%


\begin{thebibliography}{28}%
\makeatletter
\providecommand \@ifxundefined [1]{%
 \@ifx{#1\undefined}
}%
\providecommand \@ifnum [1]{%
 \ifnum #1\expandafter \@firstoftwo
 \else \expandafter \@secondoftwo
 \fi
}%
\providecommand \@ifx [1]{%
 \ifx #1\expandafter \@firstoftwo
 \else \expandafter \@secondoftwo
 \fi
}%
\providecommand \natexlab [1]{#1}%
\providecommand \enquote  [1]{``#1''}%
\providecommand \bibnamefont  [1]{#1}%
\providecommand \bibfnamefont [1]{#1}%
\providecommand \citenamefont [1]{#1}%
\providecommand \href@noop [0]{\@secondoftwo}%
\providecommand \href [0]{\begingroup \@sanitize@url \@href}%
\providecommand \@href[1]{\@@startlink{#1}\@@href}%
\providecommand \@@href[1]{\endgroup#1\@@endlink}%
\providecommand \@sanitize@url [0]{\catcode `\\12\catcode `\$12\catcode
  `\&12\catcode `\#12\catcode `\^12\catcode `\_12\catcode `\%12\relax}%
\providecommand \@@startlink[1]{}%
\providecommand \@@endlink[0]{}%
\providecommand \url  [0]{\begingroup\@sanitize@url \@url }%
\providecommand \@url [1]{\endgroup\@href {#1}{\urlprefix }}%
\providecommand \urlprefix  [0]{URL }%
\providecommand \Eprint [0]{\href }%
\providecommand \doibase [0]{http://dx.doi.org/}%
\providecommand \selectlanguage [0]{\@gobble}%
\providecommand \bibinfo  [0]{\@secondoftwo}%
\providecommand \bibfield  [0]{\@secondoftwo}%
\providecommand \translation [1]{[#1]}%
\providecommand \BibitemOpen [0]{}%
\providecommand \bibitemStop [0]{}%
\providecommand \bibitemNoStop [0]{.\EOS\space}%
\providecommand \EOS [0]{\spacefactor3000\relax}%
\providecommand \BibitemShut  [1]{\csname bibitem#1\endcsname}%
\let\auto@bib@innerbib\@empty
\bibitem [{\citenamefont {Meitner}\ and\ \citenamefont
  {Kösters}(1933)}]{meitner_uber_1933}%
  \BibitemOpen
  \bibfield  {author} {\bibinfo {author} {\bibfnamefont {L.}~\bibnamefont
  {Meitner}}\ and\ \bibinfo {author} {\bibfnamefont {H.}~\bibnamefont
  {Kösters}},\ }\bibfield  {title} {\enquote {\bibinfo {title} {Über die
  {Streuung} kurzwelliger $\gamma$-{Strahlen}},}\ }\href {\doibase
  10.1007/BF01333827} {\bibfield  {journal} {\bibinfo  {journal} {Zeitschrift
  für Physik}\ }\textbf {\bibinfo {volume} {84}},\ \bibinfo {pages} {137--144}
  (\bibinfo {year} {1933})}\BibitemShut {NoStop}%
\bibitem [{\citenamefont {Schumacher}(1999)}]{SCHUMACHER1999101}%
  \BibitemOpen
  \bibfield  {author} {\bibinfo {author} {\bibfnamefont {M.}~\bibnamefont
  {Schumacher}},\ }\bibfield  {title} {\enquote {\bibinfo {title} {Delbrück
  scattering},}\ }\href {\doibase
  https://doi.org/10.1016/S0969-806X(99)00289-3} {\bibfield  {journal}
  {\bibinfo  {journal} {Radiation Physics and Chemistry}\ }\textbf {\bibinfo
  {volume} {56}},\ \bibinfo {pages} {101--111} (\bibinfo {year}
  {1999})}\BibitemShut {NoStop}%
\bibitem [{\citenamefont {Milstein}\ and\ \citenamefont
  {Schumacher}(1994)}]{MILSTEIN1994183}%
  \BibitemOpen
  \bibfield  {author} {\bibinfo {author} {\bibfnamefont {A.I.}\ \bibnamefont
  {Milstein}}\ and\ \bibinfo {author} {\bibfnamefont {M.}~\bibnamefont
  {Schumacher}},\ }\bibfield  {title} {\enquote {\bibinfo {title} {Present
  status of {D}elbrück scattering},}\ }\href {\doibase
  https://doi.org/10.1016/0370-1573(94)00058-1} {\bibfield  {journal} {\bibinfo
   {journal} {Physics Reports}\ }\textbf {\bibinfo {volume} {243}},\ \bibinfo
  {pages} {183--214} (\bibinfo {year} {1994})}\BibitemShut {NoStop}%
\bibitem [{\citenamefont {Papatzacos}\ and\ \citenamefont
  {Mork}(1975)}]{PhysRevD.12.206}%
  \BibitemOpen
  \bibfield  {author} {\bibinfo {author} {\bibfnamefont {P.}~\bibnamefont
  {Papatzacos}}\ and\ \bibinfo {author} {\bibfnamefont {K.}~\bibnamefont
  {Mork}},\ }\bibfield  {title} {\enquote {\bibinfo {title} {{D}elbr\"uck
  scattering calculations},}\ }\href {\doibase 10.1103/PhysRevD.12.206}
  {\bibfield  {journal} {\bibinfo  {journal} {Phys. Rev. D}\ }\textbf {\bibinfo
  {volume} {12}},\ \bibinfo {pages} {206--218} (\bibinfo {year}
  {1975})}\BibitemShut {NoStop}%
\bibitem [{\citenamefont {Hartemann}\ \emph {et~al.}(2005)\citenamefont
  {Hartemann}, \citenamefont {Brown}, \citenamefont {Gibson}, \citenamefont
  {Anderson}, \citenamefont {Tremaine}, \citenamefont {Springer}, \citenamefont
  {Wootton}, \citenamefont {Hartouni},\ and\ \citenamefont
  {Barty}}]{PhysRevSTAB.8.100702}%
  \BibitemOpen
  \bibfield  {author} {\bibinfo {author} {\bibfnamefont {F.~V.}\ \bibnamefont
  {Hartemann}}, \bibinfo {author} {\bibfnamefont {W.~J.}\ \bibnamefont
  {Brown}}, \bibinfo {author} {\bibfnamefont {D.~J.}\ \bibnamefont {Gibson}},
  \bibinfo {author} {\bibfnamefont {S.~G.}\ \bibnamefont {Anderson}}, \bibinfo
  {author} {\bibfnamefont {A.~M.}\ \bibnamefont {Tremaine}}, \bibinfo {author}
  {\bibfnamefont {P.~T.}\ \bibnamefont {Springer}}, \bibinfo {author}
  {\bibfnamefont {A.~J.}\ \bibnamefont {Wootton}}, \bibinfo {author}
  {\bibfnamefont {E.~P.}\ \bibnamefont {Hartouni}}, \ and\ \bibinfo {author}
  {\bibfnamefont {C.~P.~J.}\ \bibnamefont {Barty}},\ }\bibfield  {title}
  {\enquote {\bibinfo {title} {High-energy scaling of compton scattering light
  sources},}\ }\href {\doibase 10.1103/PhysRevSTAB.8.100702} {\bibfield
  {journal} {\bibinfo  {journal} {Phys. Rev. ST Accel. Beams}\ }\textbf
  {\bibinfo {volume} {8}},\ \bibinfo {pages} {100702} (\bibinfo {year}
  {2005})}\BibitemShut {NoStop}%
\bibitem [{\citenamefont {Dupraz}\ \emph {et~al.}(2014)\citenamefont {Dupraz},
  \citenamefont {Cassou}, \citenamefont {Delerue}, \citenamefont {Fichot},
  \citenamefont {Martens}, \citenamefont {Stocchi}, \citenamefont {Variola},
  \citenamefont {Zomer}, \citenamefont {Courjaud}, \citenamefont {Mottay},
  \citenamefont {Druon}, \citenamefont {Gatti}, \citenamefont {Ghigo},
  \citenamefont {Hovsepian}, \citenamefont {Riou}, \citenamefont {Wang},
  \citenamefont {Mueller}, \citenamefont {Palumbo}, \citenamefont {Serafini},\
  and\ \citenamefont {Tomassini}}]{PhysRevSTAB.17.033501}%
  \BibitemOpen
  \bibfield  {author} {\bibinfo {author} {\bibfnamefont {K.}~\bibnamefont
  {Dupraz}}, \bibinfo {author} {\bibfnamefont {K.}~\bibnamefont {Cassou}},
  \bibinfo {author} {\bibfnamefont {N.}~\bibnamefont {Delerue}}, \bibinfo
  {author} {\bibfnamefont {P.}~\bibnamefont {Fichot}}, \bibinfo {author}
  {\bibfnamefont {A.}~\bibnamefont {Martens}}, \bibinfo {author} {\bibfnamefont
  {A.}~\bibnamefont {Stocchi}}, \bibinfo {author} {\bibfnamefont
  {A.}~\bibnamefont {Variola}}, \bibinfo {author} {\bibfnamefont
  {F.}~\bibnamefont {Zomer}}, \bibinfo {author} {\bibfnamefont
  {A.}~\bibnamefont {Courjaud}}, \bibinfo {author} {\bibfnamefont
  {E.}~\bibnamefont {Mottay}}, \bibinfo {author} {\bibfnamefont
  {F.}~\bibnamefont {Druon}}, \bibinfo {author} {\bibfnamefont
  {G.}~\bibnamefont {Gatti}}, \bibinfo {author} {\bibfnamefont
  {A.}~\bibnamefont {Ghigo}}, \bibinfo {author} {\bibfnamefont
  {T.}~\bibnamefont {Hovsepian}}, \bibinfo {author} {\bibfnamefont {J.~Y.}\
  \bibnamefont {Riou}}, \bibinfo {author} {\bibfnamefont {F.}~\bibnamefont
  {Wang}}, \bibinfo {author} {\bibfnamefont {A.~C.}\ \bibnamefont {Mueller}},
  \bibinfo {author} {\bibfnamefont {L.}~\bibnamefont {Palumbo}}, \bibinfo
  {author} {\bibfnamefont {L.}~\bibnamefont {Serafini}}, \ and\ \bibinfo
  {author} {\bibfnamefont {P.}~\bibnamefont {Tomassini}},\ }\bibfield  {title}
  {\enquote {\bibinfo {title} {Design and optimization of a highly efficient
  optical multipass system for $\ensuremath{\gamma}$-ray beam production from
  electron laser beam compton scattering},}\ }\href {\doibase
  10.1103/PhysRevSTAB.17.033501} {\bibfield  {journal} {\bibinfo  {journal}
  {Phys. Rev. ST Accel. Beams}\ }\textbf {\bibinfo {volume} {17}},\ \bibinfo
  {pages} {033501} (\bibinfo {year} {2014})}\BibitemShut {NoStop}%
\bibitem [{\citenamefont {Litvinenko}\ \emph {et~al.}(1997)\citenamefont
  {Litvinenko}, \citenamefont {Burnham}, \citenamefont {Emamian}, \citenamefont
  {Hower}, \citenamefont {Madey}, \citenamefont {Morcombe}, \citenamefont
  {O'Shea}, \citenamefont {Park}, \citenamefont {Sachtschale}, \citenamefont
  {Straub}, \citenamefont {Swift}, \citenamefont {Wang}, \citenamefont {Wu},
  \citenamefont {Canon}, \citenamefont {Howell}, \citenamefont {Roberson},
  \citenamefont {Schreiber}, \citenamefont {Spraker}, \citenamefont {Tornow},
  \citenamefont {Weller}, \citenamefont {Pinayev}, \citenamefont {Gavrilov},
  \citenamefont {Fedotov}, \citenamefont {Kulipanov}, \citenamefont {Kurkin},
  \citenamefont {Mikhailov}, \citenamefont {Popik}, \citenamefont {Skrinsky},
  \citenamefont {Vinokurov}, \citenamefont {Norum}, \citenamefont {Lumpkin},\
  and\ \citenamefont {Yang}}]{PhysRevLett.78.4569}%
  \BibitemOpen
  \bibfield  {author} {\bibinfo {author} {\bibfnamefont {V.~N.}\ \bibnamefont
  {Litvinenko}}, \bibinfo {author} {\bibfnamefont {B.}~\bibnamefont {Burnham}},
  \bibinfo {author} {\bibfnamefont {M.}~\bibnamefont {Emamian}}, \bibinfo
  {author} {\bibfnamefont {N.}~\bibnamefont {Hower}}, \bibinfo {author}
  {\bibfnamefont {J.~M.~J.}\ \bibnamefont {Madey}}, \bibinfo {author}
  {\bibfnamefont {P.}~\bibnamefont {Morcombe}}, \bibinfo {author}
  {\bibfnamefont {P.~G.}\ \bibnamefont {O'Shea}}, \bibinfo {author}
  {\bibfnamefont {S.~H.}\ \bibnamefont {Park}}, \bibinfo {author}
  {\bibfnamefont {R.}~\bibnamefont {Sachtschale}}, \bibinfo {author}
  {\bibfnamefont {K.~D.}\ \bibnamefont {Straub}}, \bibinfo {author}
  {\bibfnamefont {G.}~\bibnamefont {Swift}}, \bibinfo {author} {\bibfnamefont
  {P.}~\bibnamefont {Wang}}, \bibinfo {author} {\bibfnamefont {Y.}~\bibnamefont
  {Wu}}, \bibinfo {author} {\bibfnamefont {R.~S.}\ \bibnamefont {Canon}},
  \bibinfo {author} {\bibfnamefont {C.~R.}\ \bibnamefont {Howell}}, \bibinfo
  {author} {\bibfnamefont {N.~R.}\ \bibnamefont {Roberson}}, \bibinfo {author}
  {\bibfnamefont {E.~C.}\ \bibnamefont {Schreiber}}, \bibinfo {author}
  {\bibfnamefont {M.}~\bibnamefont {Spraker}}, \bibinfo {author} {\bibfnamefont
  {W.}~\bibnamefont {Tornow}}, \bibinfo {author} {\bibfnamefont {H.~R.}\
  \bibnamefont {Weller}}, \bibinfo {author} {\bibfnamefont {I.~V.}\
  \bibnamefont {Pinayev}}, \bibinfo {author} {\bibfnamefont {N.~G.}\
  \bibnamefont {Gavrilov}}, \bibinfo {author} {\bibfnamefont {M.~G.}\
  \bibnamefont {Fedotov}}, \bibinfo {author} {\bibfnamefont {G.~N.}\
  \bibnamefont {Kulipanov}}, \bibinfo {author} {\bibfnamefont {G.~Y.}\
  \bibnamefont {Kurkin}}, \bibinfo {author} {\bibfnamefont {S.~F.}\
  \bibnamefont {Mikhailov}}, \bibinfo {author} {\bibfnamefont {V.~M.}\
  \bibnamefont {Popik}}, \bibinfo {author} {\bibfnamefont {A.~N.}\ \bibnamefont
  {Skrinsky}}, \bibinfo {author} {\bibfnamefont {N.~A.}\ \bibnamefont
  {Vinokurov}}, \bibinfo {author} {\bibfnamefont {B.~E.}\ \bibnamefont
  {Norum}}, \bibinfo {author} {\bibfnamefont {A.}~\bibnamefont {Lumpkin}}, \
  and\ \bibinfo {author} {\bibfnamefont {B.}~\bibnamefont {Yang}},\ }\bibfield
  {title} {\enquote {\bibinfo {title} {Gamma-ray production in a storage ring
  free-electron laser},}\ }\href {\doibase 10.1103/PhysRevLett.78.4569}
  {\bibfield  {journal} {\bibinfo  {journal} {Phys. Rev. Lett.}\ }\textbf
  {\bibinfo {volume} {78}},\ \bibinfo {pages} {4569--4572} (\bibinfo {year}
  {1997})}\BibitemShut {NoStop}%
\bibitem [{\citenamefont {Albert}\ \emph {et~al.}(2011)\citenamefont {Albert},
  \citenamefont {Anderson}, \citenamefont {Gibson}, \citenamefont {Marsh},
  \citenamefont {Wu}, \citenamefont {Siders}, \citenamefont {Barty},\ and\
  \citenamefont {Hartemann}}]{PhysRevSTAB.14.050703}%
  \BibitemOpen
  \bibfield  {author} {\bibinfo {author} {\bibfnamefont {F.}~\bibnamefont
  {Albert}}, \bibinfo {author} {\bibfnamefont {S.~G.}\ \bibnamefont
  {Anderson}}, \bibinfo {author} {\bibfnamefont {D.~J.}\ \bibnamefont
  {Gibson}}, \bibinfo {author} {\bibfnamefont {R.~A.}\ \bibnamefont {Marsh}},
  \bibinfo {author} {\bibfnamefont {S.~S.}\ \bibnamefont {Wu}}, \bibinfo
  {author} {\bibfnamefont {C.~W.}\ \bibnamefont {Siders}}, \bibinfo {author}
  {\bibfnamefont {C.~P.~J.}\ \bibnamefont {Barty}}, \ and\ \bibinfo {author}
  {\bibfnamefont {F.~V.}\ \bibnamefont {Hartemann}},\ }\bibfield  {title}
  {\enquote {\bibinfo {title} {Design of narrow-band compton scattering sources
  for nuclear resonance fluorescence},}\ }\href {\doibase
  10.1103/PhysRevSTAB.14.050703} {\bibfield  {journal} {\bibinfo  {journal}
  {Phys. Rev. ST Accel. Beams}\ }\textbf {\bibinfo {volume} {14}},\ \bibinfo
  {pages} {050703} (\bibinfo {year} {2011})}\BibitemShut {NoStop}%
\bibitem [{\citenamefont {Krasny}(2015)}]{krasny_gamma_2015}%
  \BibitemOpen
  \bibfield  {author} {\bibinfo {author} {\bibfnamefont {Mieczyslaw~Witold}\
  \bibnamefont {Krasny}},\ }\href {\doibase 10.48550/arXiv.1511.07794}
  {\enquote {\bibinfo {title} {The {Gamma} {Factory} proposal for {CERN}},}\ }
  (\bibinfo {year} {2015}),\ \bibinfo {note} {arXiv:1511.07794 [hep-ex,
  physics:hep-ph, physics:physics]}\BibitemShut {NoStop}%
\bibitem [{\citenamefont {Budker}\ \emph {et~al.}(2020)\citenamefont {Budker},
  \citenamefont {Crespo L\'{o}pez-Urrutia}, \citenamefont {Derevianko},
  \citenamefont {Flambaum}, \citenamefont {Krasny}, \citenamefont {Petrenko},
  \citenamefont {Pustelny}, \citenamefont {Surzhykov}, \citenamefont
  {Yerokhin},\ and\ \citenamefont {Zolotorev}}]{budker_atomic_2020}%
  \BibitemOpen
  \bibfield  {author} {\bibinfo {author} {\bibfnamefont {Dmitry}\ \bibnamefont
  {Budker}}, \bibinfo {author} {\bibfnamefont {Jos\'{e}~R.}\ \bibnamefont
  {Crespo L\'{o}pez-Urrutia}}, \bibinfo {author} {\bibfnamefont {Andrei}\
  \bibnamefont {Derevianko}}, \bibinfo {author} {\bibfnamefont {Victor~V.}\
  \bibnamefont {Flambaum}}, \bibinfo {author} {\bibfnamefont
  {Mieczyslaw~Witold}\ \bibnamefont {Krasny}}, \bibinfo {author} {\bibfnamefont
  {Alexey}\ \bibnamefont {Petrenko}}, \bibinfo {author} {\bibfnamefont
  {Szymon}\ \bibnamefont {Pustelny}}, \bibinfo {author} {\bibfnamefont
  {Andrey}\ \bibnamefont {Surzhykov}}, \bibinfo {author} {\bibfnamefont
  {Vladimir~A.}\ \bibnamefont {Yerokhin}}, \ and\ \bibinfo {author}
  {\bibfnamefont {Max}\ \bibnamefont {Zolotorev}},\ }\bibfield  {title}
  {\enquote {\bibinfo {title} {Atomic {Physics} {Studies} at the {Gamma}
  {Factory} at {CERN}},}\ }\href {\doibase 10.1002/andp.202000204} {\bibfield
  {journal} {\bibinfo  {journal} {Annalen der Physik}\ }\textbf {\bibinfo
  {volume} {532}},\ \bibinfo {pages} {2000204} (\bibinfo {year}
  {2020})}\BibitemShut {NoStop}%
\bibitem [{\citenamefont {Rullhusen}\ \emph
  {et~al.}(1979{\natexlab{a}})\citenamefont {Rullhusen}, \citenamefont
  {Smend},\ and\ \citenamefont {Schumacher}}]{RULLHUSEN1979166}%
  \BibitemOpen
  \bibfield  {author} {\bibinfo {author} {\bibfnamefont {P.}~\bibnamefont
  {Rullhusen}}, \bibinfo {author} {\bibfnamefont {F.}~\bibnamefont {Smend}}, \
  and\ \bibinfo {author} {\bibfnamefont {M.}~\bibnamefont {Schumacher}},\
  }\bibfield  {title} {\enquote {\bibinfo {title} {Delbrück scattering of 2754
  kev photons by {Nd, Ce, I, Sn, Mo and Zn}},}\ }\href {\doibase
  https://doi.org/10.1016/0370-2693(79)90274-0} {\bibfield  {journal} {\bibinfo
   {journal} {Physics Letters B}\ }\textbf {\bibinfo {volume} {84}},\ \bibinfo
  {pages} {166--168} (\bibinfo {year} {1979}{\natexlab{a}})}\BibitemShut
  {NoStop}%
\bibitem [{\citenamefont {Rullhusen}\ \emph
  {et~al.}(1979{\natexlab{b}})\citenamefont {Rullhusen}, \citenamefont {Smend},
  \citenamefont {Schumacher}, \citenamefont {Hanser},\ and\ \citenamefont
  {Rebel}}]{rullhusen_coulomb_1979}%
  \BibitemOpen
  \bibfield  {author} {\bibinfo {author} {\bibfnamefont {P.}~\bibnamefont
  {Rullhusen}}, \bibinfo {author} {\bibfnamefont {F.}~\bibnamefont {Smend}},
  \bibinfo {author} {\bibfnamefont {M.}~\bibnamefont {Schumacher}}, \bibinfo
  {author} {\bibfnamefont {A.}~\bibnamefont {Hanser}}, \ and\ \bibinfo {author}
  {\bibfnamefont {H.}~\bibnamefont {Rebel}},\ }\bibfield  {title} {\enquote
  {\bibinfo {title} {Coulomb correction to {Delbrück} scattering investigated
  at {Z=94}},}\ }\href {\doibase 10.1007/BF01435270} {\bibfield  {journal}
  {\bibinfo  {journal} {Zeitschrift für Physik A Atoms and Nuclei}\ }\textbf
  {\bibinfo {volume} {293}},\ \bibinfo {pages} {287--292} (\bibinfo {year}
  {1979}{\natexlab{b}})}\BibitemShut {NoStop}%
\bibitem [{\citenamefont {Rullhusen}\ \emph {et~al.}(1981)\citenamefont
  {Rullhusen}, \citenamefont {M\"uckenheim}, \citenamefont {Smend},
  \citenamefont {Schumacher}, \citenamefont {Berg}, \citenamefont {Mork},\ and\
  \citenamefont {Kissel}}]{PhysRevC.23.1375}%
  \BibitemOpen
  \bibfield  {author} {\bibinfo {author} {\bibfnamefont {P.}~\bibnamefont
  {Rullhusen}}, \bibinfo {author} {\bibfnamefont {W.}~\bibnamefont
  {M\"uckenheim}}, \bibinfo {author} {\bibfnamefont {F.}~\bibnamefont {Smend}},
  \bibinfo {author} {\bibfnamefont {M.}~\bibnamefont {Schumacher}}, \bibinfo
  {author} {\bibfnamefont {G.~P.~A.}\ \bibnamefont {Berg}}, \bibinfo {author}
  {\bibfnamefont {K.}~\bibnamefont {Mork}}, \ and\ \bibinfo {author}
  {\bibfnamefont {Lynn}\ \bibnamefont {Kissel}},\ }\bibfield  {title} {\enquote
  {\bibinfo {title} {Test of vacuum polarization by precise investigation of
  {D}elbr\"uck scattering},}\ }\href {\doibase 10.1103/PhysRevC.23.1375}
  {\bibfield  {journal} {\bibinfo  {journal} {Phys. Rev. C}\ }\textbf {\bibinfo
  {volume} {23}},\ \bibinfo {pages} {1375--1383} (\bibinfo {year}
  {1981})}\BibitemShut {NoStop}%
\bibitem [{\citenamefont {Schumacher}\ \emph {et~al.}(1975)\citenamefont
  {Schumacher}, \citenamefont {Borchert}, \citenamefont {Smend},\ and\
  \citenamefont {Rullhusen}}]{SCHUMACHER1975134}%
  \BibitemOpen
  \bibfield  {author} {\bibinfo {author} {\bibfnamefont {M.}~\bibnamefont
  {Schumacher}}, \bibinfo {author} {\bibfnamefont {I.}~\bibnamefont
  {Borchert}}, \bibinfo {author} {\bibfnamefont {F.}~\bibnamefont {Smend}}, \
  and\ \bibinfo {author} {\bibfnamefont {P.}~\bibnamefont {Rullhusen}},\
  }\bibfield  {title} {\enquote {\bibinfo {title} {Delbrück scattering of 2.75
  {MeV} photons by lead},}\ }\href {\doibase
  https://doi.org/10.1016/0370-2693(75)90685-1} {\bibfield  {journal} {\bibinfo
   {journal} {Physics Letters B}\ }\textbf {\bibinfo {volume} {59}},\ \bibinfo
  {pages} {134--136} (\bibinfo {year} {1975})}\BibitemShut {NoStop}%
\bibitem [{\citenamefont {Scherdin}\ \emph {et~al.}(1992)\citenamefont
  {Scherdin}, \citenamefont {Sch\"afer}, \citenamefont {Greiner},\ and\
  \citenamefont {Soff}}]{PhysRevD.45.2982}%
  \BibitemOpen
  \bibfield  {author} {\bibinfo {author} {\bibfnamefont {A.}~\bibnamefont
  {Scherdin}}, \bibinfo {author} {\bibfnamefont {A.}~\bibnamefont {Sch\"afer}},
  \bibinfo {author} {\bibfnamefont {W.}~\bibnamefont {Greiner}}, \ and\
  \bibinfo {author} {\bibfnamefont {G.}~\bibnamefont {Soff}},\ }\bibfield
  {title} {\enquote {\bibinfo {title} {{D}elbr\"uck scattering in a strong
  external field},}\ }\href {\doibase 10.1103/PhysRevD.45.2982} {\bibfield
  {journal} {\bibinfo  {journal} {Phys. Rev. D}\ }\textbf {\bibinfo {volume}
  {45}},\ \bibinfo {pages} {2982--2987} (\bibinfo {year} {1992})}\BibitemShut
  {NoStop}%
\bibitem [{\citenamefont {Scherdin}\ \emph {et~al.}(1995)\citenamefont
  {Scherdin}, \citenamefont {Schäfer}, \citenamefont {Greiner}, \citenamefont
  {Soff},\ and\ \citenamefont {Mohr}}]{scherdin_coulomb_1995}%
  \BibitemOpen
  \bibfield  {author} {\bibinfo {author} {\bibfnamefont {A.}~\bibnamefont
  {Scherdin}}, \bibinfo {author} {\bibfnamefont {A.}~\bibnamefont {Schäfer}},
  \bibinfo {author} {\bibfnamefont {W.}~\bibnamefont {Greiner}}, \bibinfo
  {author} {\bibfnamefont {G.}~\bibnamefont {Soff}}, \ and\ \bibinfo {author}
  {\bibfnamefont {P.~J.}\ \bibnamefont {Mohr}},\ }\bibfield  {title} {\enquote
  {\bibinfo {title} {Coulomb corrections to {Delbrück} scattering},}\ }\href
  {\doibase 10.1007/BF01292332} {\bibfield  {journal} {\bibinfo  {journal}
  {Zeitschrift für Physik A Hadrons and Nuclei}\ }\textbf {\bibinfo {volume}
  {353}},\ \bibinfo {pages} {273--277} (\bibinfo {year} {1995})}\BibitemShut
  {NoStop}%
\bibitem [{\citenamefont {Sommerfeldt}\ \emph {et~al.}(2022)\citenamefont
  {Sommerfeldt}, \citenamefont {Yerokhin}, \citenamefont {M\"uller},
  \citenamefont {Zaytsev}, \citenamefont {Volotka},\ and\ \citenamefont
  {Surzhykov}}]{PhysRevA.105.022804}%
  \BibitemOpen
  \bibfield  {author} {\bibinfo {author} {\bibfnamefont {J.}~\bibnamefont
  {Sommerfeldt}}, \bibinfo {author} {\bibfnamefont {V.~A.}\ \bibnamefont
  {Yerokhin}}, \bibinfo {author} {\bibfnamefont {R.~A.}\ \bibnamefont
  {M\"uller}}, \bibinfo {author} {\bibfnamefont {V.~A.}\ \bibnamefont
  {Zaytsev}}, \bibinfo {author} {\bibfnamefont {A.~V.}\ \bibnamefont
  {Volotka}}, \ and\ \bibinfo {author} {\bibfnamefont {A.}~\bibnamefont
  {Surzhykov}},\ }\bibfield  {title} {\enquote {\bibinfo {title} {Calculations
  of {D}elbr\"uck scattering to all orders in $\ensuremath{\alpha}{Z}$},}\
  }\href {\doibase 10.1103/PhysRevA.105.022804} {\bibfield  {journal} {\bibinfo
   {journal} {Phys. Rev. A}\ }\textbf {\bibinfo {volume} {105}},\ \bibinfo
  {pages} {022804} (\bibinfo {year} {2022})}\BibitemShut {NoStop}%
\bibitem [{\citenamefont {Mohr}\ \emph {et~al.}(1998)\citenamefont {Mohr},
  \citenamefont {Plunien},\ and\ \citenamefont {Soff}}]{MOHR1998227}%
  \BibitemOpen
  \bibfield  {author} {\bibinfo {author} {\bibfnamefont {P.~J.}\ \bibnamefont
  {Mohr}}, \bibinfo {author} {\bibfnamefont {G.}~\bibnamefont {Plunien}}, \
  and\ \bibinfo {author} {\bibfnamefont {G.}~\bibnamefont {Soff}},\ }\bibfield
  {title} {\enquote {\bibinfo {title} {{QED} corrections in heavy atoms},}\
  }\href {\doibase https://doi.org/10.1016/S0370-1573(97)00046-X} {\bibfield
  {journal} {\bibinfo  {journal} {Physics Reports}\ }\textbf {\bibinfo {volume}
  {293}},\ \bibinfo {pages} {227--369} (\bibinfo {year} {1998})}\BibitemShut
  {NoStop}%
\bibitem [{\citenamefont {Shabaev}(2002)}]{shabaev_two-time_2002}%
  \BibitemOpen
  \bibfield  {author} {\bibinfo {author} {\bibfnamefont {V.~M.}\ \bibnamefont
  {Shabaev}},\ }\bibfield  {title} {\enquote {\bibinfo {title} {Two-time
  {Green}'s function method in quantum electrodynamics of high-{Z} few-electron
  atoms},}\ }\href {\doibase 10.1016/S0370-1573(01)00024-2} {\bibfield
  {journal} {\bibinfo  {journal} {Physics Reports}\ }\textbf {\bibinfo {volume}
  {356}},\ \bibinfo {pages} {119--228} (\bibinfo {year} {2002})}\BibitemShut
  {NoStop}%
\bibitem [{\citenamefont {Soguel}\ \emph
  {et~al.}(2021{\natexlab{a}})\citenamefont {Soguel}, \citenamefont {Volotka},
  \citenamefont {Glazov},\ and\ \citenamefont {Fritzsche}}]{sym13061014}%
  \BibitemOpen
  \bibfield  {author} {\bibinfo {author} {\bibfnamefont {R.~N.}\ \bibnamefont
  {Soguel}}, \bibinfo {author} {\bibfnamefont {A.~V.}\ \bibnamefont {Volotka}},
  \bibinfo {author} {\bibfnamefont {D.~A.}\ \bibnamefont {Glazov}}, \ and\
  \bibinfo {author} {\bibfnamefont {S.}~\bibnamefont {Fritzsche}},\ }\bibfield
  {title} {\enquote {\bibinfo {title} {Many-electron {QED} with redefined
  vacuum approach},}\ }\href {https://www.mdpi.com/2073-8994/13/6/1014}
  {\bibfield  {journal} {\bibinfo  {journal} {Symmetry}\ }\textbf {\bibinfo
  {volume} {13}} (\bibinfo {year} {2021}{\natexlab{a}})}\BibitemShut {NoStop}%
\bibitem [{\citenamefont {Soguel}\ \emph
  {et~al.}(2021{\natexlab{b}})\citenamefont {Soguel}, \citenamefont {Volotka},
  \citenamefont {Tryapitsyna}, \citenamefont {Glazov}, \citenamefont
  {Kosheleva},\ and\ \citenamefont {Fritzsche}}]{PhysRevA.103.042818}%
  \BibitemOpen
  \bibfield  {author} {\bibinfo {author} {\bibfnamefont {R.~N.}\ \bibnamefont
  {Soguel}}, \bibinfo {author} {\bibfnamefont {A.~V.}\ \bibnamefont {Volotka}},
  \bibinfo {author} {\bibfnamefont {E.~V.}\ \bibnamefont {Tryapitsyna}},
  \bibinfo {author} {\bibfnamefont {D.~A.}\ \bibnamefont {Glazov}}, \bibinfo
  {author} {\bibfnamefont {V.~P.}\ \bibnamefont {Kosheleva}}, \ and\ \bibinfo
  {author} {\bibfnamefont {S.}~\bibnamefont {Fritzsche}},\ }\bibfield  {title}
  {\enquote {\bibinfo {title} {Redefined vacuum approach and gauge-invariant
  subsets in two-photon-exchange diagrams for a closed-shell system with a
  valence electron},}\ }\href {\doibase 10.1103/PhysRevA.103.042818} {\bibfield
   {journal} {\bibinfo  {journal} {Phys. Rev. A}\ }\textbf {\bibinfo {volume}
  {103}},\ \bibinfo {pages} {042818} (\bibinfo {year}
  {2021}{\natexlab{b}})}\BibitemShut {NoStop}%
\bibitem [{\citenamefont {Surzhykov}\ \emph {et~al.}(2015)\citenamefont
  {Surzhykov}, \citenamefont {Yerokhin}, \citenamefont {Stöhlker},\ and\
  \citenamefont {Fritzsche}}]{Surzhykov_2015}%
  \BibitemOpen
  \bibfield  {author} {\bibinfo {author} {\bibfnamefont {A.}~\bibnamefont
  {Surzhykov}}, \bibinfo {author} {\bibfnamefont {V.~A.}\ \bibnamefont
  {Yerokhin}}, \bibinfo {author} {\bibfnamefont {Th.}\ \bibnamefont
  {Stöhlker}}, \ and\ \bibinfo {author} {\bibfnamefont {S.}~\bibnamefont
  {Fritzsche}},\ }\bibfield  {title} {\enquote {\bibinfo {title} {Rayleigh
  x-ray scattering from many-electron atoms and ions},}\ }\href {\doibase
  10.1088/0953-4075/48/14/144015} {\bibfield  {journal} {\bibinfo  {journal}
  {Journal of Physics B: Atomic, Molecular and Optical Physics}\ }\textbf
  {\bibinfo {volume} {48}},\ \bibinfo {pages} {144015} (\bibinfo {year}
  {2015})}\BibitemShut {NoStop}%
\bibitem [{\citenamefont {Falkenberg}\ \emph {et~al.}(1992)\citenamefont
  {Falkenberg}, \citenamefont {Hünger}, \citenamefont {Rullhusen},
  \citenamefont {Schumacher}, \citenamefont {Milstein},\ and\ \citenamefont
  {Mork}}]{FALKENBERG19921}%
  \BibitemOpen
  \bibfield  {author} {\bibinfo {author} {\bibfnamefont {H.}~\bibnamefont
  {Falkenberg}}, \bibinfo {author} {\bibfnamefont {A.}~\bibnamefont {Hünger}},
  \bibinfo {author} {\bibfnamefont {P.}~\bibnamefont {Rullhusen}}, \bibinfo
  {author} {\bibfnamefont {M.}~\bibnamefont {Schumacher}}, \bibinfo {author}
  {\bibfnamefont {A.I.}\ \bibnamefont {Milstein}}, \ and\ \bibinfo {author}
  {\bibfnamefont {K.}~\bibnamefont {Mork}},\ }\bibfield  {title} {\enquote
  {\bibinfo {title} {Amplitudes for {D}elbrück scattering},}\ }\href {\doibase
  https://doi.org/10.1016/0092-640X(92)90023-B} {\bibfield  {journal} {\bibinfo
   {journal} {Atomic Data and Nuclear Data Tables}\ }\textbf {\bibinfo {volume}
  {50}},\ \bibinfo {pages} {1--27} (\bibinfo {year} {1992})}\BibitemShut
  {NoStop}%
\bibitem [{\citenamefont {Hütt}\ \emph {et~al.}(2000)\citenamefont {Hütt},
  \citenamefont {L'vov}, \citenamefont {Milstein},\ and\ \citenamefont
  {Schumacher}}]{HUTT2000457}%
  \BibitemOpen
  \bibfield  {author} {\bibinfo {author} {\bibfnamefont {M.-Th.}\ \bibnamefont
  {Hütt}}, \bibinfo {author} {\bibfnamefont {A.I.}\ \bibnamefont {L'vov}},
  \bibinfo {author} {\bibfnamefont {A.I.}\ \bibnamefont {Milstein}}, \ and\
  \bibinfo {author} {\bibfnamefont {M.}~\bibnamefont {Schumacher}},\ }\bibfield
   {title} {\enquote {\bibinfo {title} {Compton scattering by nuclei},}\ }\href
  {\doibase https://doi.org/10.1016/S0370-1573(99)00041-1} {\bibfield
  {journal} {\bibinfo  {journal} {Physics Reports}\ }\textbf {\bibinfo {volume}
  {323}},\ \bibinfo {pages} {457--594} (\bibinfo {year} {2000})}\BibitemShut
  {NoStop}%
\bibitem [{\citenamefont {Low}(1954)}]{PhysRev.96.1428}%
  \BibitemOpen
  \bibfield  {author} {\bibinfo {author} {\bibfnamefont {F.~E.}\ \bibnamefont
  {Low}},\ }\bibfield  {title} {\enquote {\bibinfo {title} {Scattering of light
  of very low frequency by systems of spin $\frac{1}{2}$},}\ }\href {\doibase
  10.1103/PhysRev.96.1428} {\bibfield  {journal} {\bibinfo  {journal} {Phys.
  Rev.}\ }\textbf {\bibinfo {volume} {96}},\ \bibinfo {pages} {1428--1432}
  (\bibinfo {year} {1954})}\BibitemShut {NoStop}%
\bibitem [{\citenamefont {Volotka}\ \emph {et~al.}(2016)\citenamefont
  {Volotka}, \citenamefont {Yerokhin}, \citenamefont {Surzhykov}, \citenamefont
  {St\"ohlker},\ and\ \citenamefont {Fritzsche}}]{PhysRevA.93.023418}%
  \BibitemOpen
  \bibfield  {author} {\bibinfo {author} {\bibfnamefont {A.~V.}\ \bibnamefont
  {Volotka}}, \bibinfo {author} {\bibfnamefont {V.~A.}\ \bibnamefont
  {Yerokhin}}, \bibinfo {author} {\bibfnamefont {A.}~\bibnamefont {Surzhykov}},
  \bibinfo {author} {\bibfnamefont {Th.}\ \bibnamefont {St\"ohlker}}, \ and\
  \bibinfo {author} {\bibfnamefont {S.}~\bibnamefont {Fritzsche}},\ }\bibfield
  {title} {\enquote {\bibinfo {title} {Many-electron effects on x-ray rayleigh
  scattering by highly charged he-like ions},}\ }\href {\doibase
  10.1103/PhysRevA.93.023418} {\bibfield  {journal} {\bibinfo  {journal} {Phys.
  Rev. A}\ }\textbf {\bibinfo {volume} {93}},\ \bibinfo {pages} {023418}
  (\bibinfo {year} {2016})}\BibitemShut {NoStop}%
\bibitem [{\citenamefont {Middents}\ \emph {et~al.}(2023)\citenamefont
  {Middents}, \citenamefont {Weber}, \citenamefont {Gumberidze}, \citenamefont
  {Hahn}, \citenamefont {Krings}, \citenamefont {Kurz}, \citenamefont
  {Pf\"afflein}, \citenamefont {Schell}, \citenamefont {Spillmann},
  \citenamefont {Strnat}, \citenamefont {Vockert}, \citenamefont {Volotka},
  \citenamefont {Surzhykov},\ and\ \citenamefont
  {St\"ohlker}}]{PhysRevA.107.012805}%
  \BibitemOpen
  \bibfield  {author} {\bibinfo {author} {\bibfnamefont {W.}~\bibnamefont
  {Middents}}, \bibinfo {author} {\bibfnamefont {G.}~\bibnamefont {Weber}},
  \bibinfo {author} {\bibfnamefont {A.}~\bibnamefont {Gumberidze}}, \bibinfo
  {author} {\bibfnamefont {C.}~\bibnamefont {Hahn}}, \bibinfo {author}
  {\bibfnamefont {T.}~\bibnamefont {Krings}}, \bibinfo {author} {\bibfnamefont
  {N.}~\bibnamefont {Kurz}}, \bibinfo {author} {\bibfnamefont {P.}~\bibnamefont
  {Pf\"afflein}}, \bibinfo {author} {\bibfnamefont {N.}~\bibnamefont {Schell}},
  \bibinfo {author} {\bibfnamefont {U.}~\bibnamefont {Spillmann}}, \bibinfo
  {author} {\bibfnamefont {S.}~\bibnamefont {Strnat}}, \bibinfo {author}
  {\bibfnamefont {M.}~\bibnamefont {Vockert}}, \bibinfo {author} {\bibfnamefont
  {A.}~\bibnamefont {Volotka}}, \bibinfo {author} {\bibfnamefont
  {A.}~\bibnamefont {Surzhykov}}, \ and\ \bibinfo {author} {\bibfnamefont
  {T.}~\bibnamefont {St\"ohlker}},\ }\bibfield  {title} {\enquote {\bibinfo
  {title} {Angle-differential cross sections for rayleigh scattering of highly
  linearly polarized hard x rays on au atoms},}\ }\href {\doibase
  10.1103/PhysRevA.107.012805} {\bibfield  {journal} {\bibinfo  {journal}
  {Phys. Rev. A}\ }\textbf {\bibinfo {volume} {107}},\ \bibinfo {pages}
  {012805} (\bibinfo {year} {2023})}\BibitemShut {NoStop}%
\bibitem [{\citenamefont {Kane}\ \emph {et~al.}(1986)\citenamefont {Kane},
  \citenamefont {Kissel}, \citenamefont {Pratt},\ and\ \citenamefont
  {Roy}}]{KANE198675}%
  \BibitemOpen
  \bibfield  {author} {\bibinfo {author} {\bibfnamefont {P.P.}\ \bibnamefont
  {Kane}}, \bibinfo {author} {\bibfnamefont {Lynn}\ \bibnamefont {Kissel}},
  \bibinfo {author} {\bibfnamefont {R.H.}\ \bibnamefont {Pratt}}, \ and\
  \bibinfo {author} {\bibfnamefont {S.C.}\ \bibnamefont {Roy}},\ }\bibfield
  {title} {\enquote {\bibinfo {title} {Elastic scattering of $\gamma$-rays and
  {X}-rays by atoms},}\ }\href {\doibase
  https://doi.org/10.1016/0370-1573(86)90018-9} {\bibfield  {journal} {\bibinfo
   {journal} {Physics Reports}\ }\textbf {\bibinfo {volume} {140}},\ \bibinfo
  {pages} {75--159} (\bibinfo {year} {1986})}\BibitemShut {NoStop}%
\end{thebibliography}
\end{document}